# On the Injunction of XAIxArt

Moving beyond explanation to sense-making

Cheshta Arora
Independent Researcher
India
cheshtaarora@outlook.in

Debarun Sarkar†
Independent Researcher
India
debarun@outlook.com

## ABSTRACT

The position paper highlights the range of concerns that are engulfed in the injunction of explainable artificial intelligence in art (XAIxArt). Through a series of quick sub-questions, it points towards the ambiguities concerning 'explanation' and the postpositivist tradition of 'relevant explanation'. Rejecting both 'explanation' and 'relevant explanation', the paper takes a stance that XAIxArt is a symptom of insecurity of the anthropocentric notion of art and a nostalgic desire to return to outmoded notions of authorship and human agency. To justify this stance, the paper makes a distinction between an ornamentation model of explanation to a model of explanation as sense-making.

## 1 Introduction

Before any discussion on AIxArt, it is necessary to delineate what AIxArt entails. Manovich [6] defines AI arts in at least three possible ways: 1) AI arts that "only simulate[s] the historical art" and "is not capable of executing the main strategy of modern art - constantly expanding what counts as art" [6:3], 2) "all methods developed in computer art since 1950s are equally valid instances of 'AI arts'" which means that "What defines whether something is "AI" is not a method but the amount and type of control we exercise over algorithmic process" [6:5] and 3) "AI art is type of art that we humans are not able to create because of the limitations of our bodies, brains, and other constraints." [6:8]

Imagine three different hypothetical exhibitions based on Manovich's typology. What would XAI look like in each case? In the first case, since the artists remain limited by AIxArt's inability to expand what counts as art, the explanation would serve to defend AI art as art. In the second case, a curatorial exhibition premised on such an approach would be deeply invested in explaining the history of computer art and situating each work along the arrow of time and explaining the amount of control on algorithmic processes. Finally, for the third exhibition, an explanation either would not be necessary (as the newness of the phenomenon would be enough to capture the consumer) or would merely point towards our limitations.

We are, however, not interested in creating a typology of exhibitions that may or may not require explanation. Rather, we are invested in understanding 'explanation', the purpose it serves and how it relates to the larger experience of encountering an artefact. That is, in one's experience of an artwork, what would be the role of explanation in curating that experience?

## 2 The injunction of XAIxArts

Explainable AI has become one of the most recent moral injunctions given the ever-increasing awareness of predictive AI systems and their accompanying harms. In the context of predictive machines, to explain is to situate the AI system into a here and now. It is to interrogate its monopoly on knowledge and truth and offer a guided tour away from "an AI system knows" to "an AI system knows *because*…". While in the philosophy of science, the problem of explanation was more of an epistemic concern (i.e., how explanations relate to the question of knowledge), in the field of AI ethics, an explainable AI is primarily an ethical concern. This shift from epistemic to ethical is grounded in the liberal, secular and rational notion of morality which exists amidst and in full recognition of the hierarchical socio-political relations. That is, a decision-maker affecting our life, *at the very least,* owes us an explanation.

Given this, what would it mean to import this injunction into the domain of arts and aesthetics? It is clear that an exploration of this question will raise further questions, some of which are listed here:

- What is the precise object of explanation?—Is it the technique, the general character of AI systems or the artefact itself?
- What purpose would it serve?—Is it to expand one's understanding of the history of arts and aesthetics? Or, to expand one's critical understanding of pervasive AI systems more broadly? Or, to justify the artefact as a work of art and its place in a gallery?
- What will the explanation look like?—Like a formal logical deduction? Or, linear, cause & effect kind of expositions? Well-informed? An ordinary explanation?
- Where will explanations be placed?—As essentially outside of the artwork? As part of the artwork? Or as escorts of the artwork?
- Who decides what needs explaining? Or who distinguishes a good explanation from a bad one?
- If explanations explain, then what do artworks do?
- Finally, what do explanations exactly do to one's encounter with the artwork?—Does it explain? Teach? Preach? Communicate? Facilitate meaning-making?

In place of a good explanation or answer to the previous question, the series of questions help demonstrate that the injunction to make AI arts explainable is thornier than it might appear. The questions, however, do allow one to see the act of explanation in broad strokes, the specificities of which have been mapped at various times by different scholars. It is more or less accepted that explanations operate in certain relations of hierarchy and that explanations are social [7] i.e., the "object of explanation… is not a simple object, like an event or a state of affairs, but more like a state of affairs together with a definite *space of alternatives"* which produce "different things to be-explained, two different objects of explanation"[5]. This consensus merely points towards the importance of taking a nuanced approach vis-à-vis explanation. While we agree that explanations are relative, subjective, and social, it is not impossible to argue in favour of and arrive at 'relevant' explanations which are more just, more conducive to human, or more-than-human welfare [5]. However even a situated understanding of explanation doesn't exactly answer the question posed previously—what would it mean to import this injunction into the domain of arts and aesthetics? The spectre of relevant explanation can sit side by side with the moral injunction of explainable AI arts. The question of import can simply be cast aside.

In this paper, however, we aim to hold on to the question and take a provocative stance that the import of XAI into arts is a symptom of insecurity of the anthropocentric notion of art which aims to reduce and reify AI art through human framings. It is a symptom of a nostalgic desire to return to outmoded notions of authorship and human agency. Some clarification regarding explanation in contemporary art is called for to situate our position.

## 3  To explain is to ornament

'To explain' is to elaborate and to extend. In art practices, this manifests as curatorial notes for example, which act as decorative ornate that so often justify and elevate an artefact to the status of art i.e., it evaluates and assigns value to the artwork. The curatorial note 'explains' the artwork which in turn re/de/values the work of art itself. It devalues the artwork in the sense that it is an acknowledgement of a lack of trust in the artwork and its form, medium and content.

Art exists in a zone or neighbourhood of 'no orientation', in a zone of non-logical reasoning. As Choudhury [2] notes, the figure of the artist is one of the rare figures which claims a lack of orientation through a lack of orientation, unlike the figure of the philosopher (with whom we can certainly include the curator), who claims a lack of orientation while having traversed a path of orientation—of logical reasoning. Why is this journey towards 'no orientation' so significant for the artist, the philosopher and the curator one may ask? It is to encounter the 'new', albeit via different routes.

If one agrees with the basic principle that art concerns itself with a certain *je sais pas quoi* i.e., the sublime, then any act of explaining or orientation is necessarily an ornamentation that is always located outside the artwork. But while being located outside it relates to the work. It devalues the artwork because of its inability to claim orientation. It simultaneously revalues it by explaining it, justifying its existence in an exhibition, gallery, what have you, noting its method in a 'scientific' manner and then situating it in art history.

This evaluative role of curatorial note reaches absurd limits when one encounters a found object in a gallery. It is not an exaggeration to suggest that every visitor to a contemporary art exhibition has had a moment of disorientation in a gallery where a found object is elevated to art through a curatorial note while the audience is left wondering if a random artefact in the gallery is part of the exhibition/installation only to realize that the curatorial note is missing. One could argue that it is precisely this disorientation that the medium of found object aims to invoke, yet this Brechtian alienation effect is a side effect which artists and curators do not intend anymore in today's glossy worlds of art galleries. The audience at that moment experiences, to put it in Badiou's words, truth as an event through the truth procedure of art wherein commodity fetishism is made evident [1]. Yet, this experience is not the intended experience of such exhibitions. The initial rebellious spirit of Dadaism is channelled through today's art circuits to construct spectacles [3]. The aim of such spectacles is not to encounter the sublime, but to plug the artist, curator, artwork and consumer into a relationship of exchange and value.

While curators today have come up with codified handbooks to aid curating exhibitions and explicitly prescribe explanations to ensure that the exhibition was a "success". At the same time, there always remains another possibility of throwing an unsuspecting audience into a relationship with an artwork. The failure to establish the 'right' relationship, and communicate the intended meaning, in such a case might not be seen as 'problem'. Consider simply the case of a linear video game that forces the player to stick to the storyline against a video game that allows meandering and re-joining storylines and leaving them without a clear tutorial.

## 4 Explanation vs. sense-making

Yet, one can argue that art itself is an act of explaining. Here, it is important to distinguish between explanation, and sense-making. In the positivist system, an explanation as one of the identified components of the theory of knowledge is a "formal deduction, a single, uniform model of a single, complete, correct explanation for a given phenomenon" [5]. In the postpositivist tradition, it is reduced to a 'relevant explanation' as opposed to a single, correct explanation. These discussions, however, are still tied to a theory of knowledge.

In the domain of art, however, we have been trying to distinguish (relevant) explanation vs sense-making—of art as a plurality and a shared experience of the sublime therein.

For the widest possibility of AIxArt, a narrow notion of sense that limits itself to "linguistic or logical, order of representation" [9:9] fails to account for the multi-modal and processual nature of AIxArt wherein the human, AI, art and the host of object relations are in a mode of becoming. Such a context necessitates a notion of sense understood as "'a system of echoes, of resumptions and resonances' [4:170/199 quoted] between series of sense-events" [9:22–23]. Sense-making's disavowal of any evaluation and control of bodies, objects, relations and meaning-making makes possible an immanent experience of art that emerges in relationality (and not in understanding). To impose a logic of explanation in such a scenario is to foreclose the very possibility of sense-making and its invitation to becoming.

Moving away from explanation to sense-making, what would a reconfigured curatorial note look like? A curatorial note in such an approach would not just explain but be an integral part of the artistic experience. Consider for example an AI which writes curatorial notes, makes errors, and conjures new unaccounted relationalities into being that transform "the spatiotemporality of experience" [8:53]. It is exigent in such a case to move away from the ornamentation model of explanation to a model of explanation as sense-making that is intimately aware of various processes' and objects' place in a network of relationships between various human and non-human actors and the power of evaluation that material-semiotic relationships operationalize i.e., it moves away from explanation to sense-making.

## FUNDING

This material is based upon work supported in whole or in part by The Notre Dame-IBM Tech Ethics Lab. Such support does not constitute endorsement by the sponsor of the views expressed in this publication.